\documentclass{article}
\usepackage{authblk}
\usepackage{geometry}
\usepackage[draft]{hyperref}
\usepackage{graphicx}
\usepackage{caption}
\usepackage{ifthen}
\usepackage{url}
\usepackage{enumitem}
\usepackage{siunitx}
\DeclareSIUnit\year{yr}
\DeclareSIUnit\MWh{MWh}

\usepackage[style=numeric, sorting=none]{biblatex}
\usepackage{tabularx} 

\DeclareCaptionLabelFormat{figlabel}{{Fig. #2. }}
\DeclareCaptionLabelFormat{tablelabel}{{Table #2. }}
\begin{document}


\renewcommand{\thefootnote}{\fnsymbol{footnote}}
\title{Valuing maintenance strategies for fusion plants as part of a future electricity grid}
\author[1]{Jacob A. Schwartz \thanks{Corresponding author: jschwartz@pppl.gov}}
\author[2,3]{W. Ricks}
\author[2,3]{E. Kolemen}
\author[2,3]{J. D. Jenkins}
\affil[1]{Princeton Plasma Physics Laboratory}
\affil[2]{Princeton University, Dept. of Mechanical and Aerospace Engineering}
\affil[3]{Princeton University, Andlinger Center for Energy and the Environment}
 
\date{}
\maketitle
\renewcommand{\thefootnote}{\arabic{footnote}}
\begin{abstract}

Scheduled maintenance is likely to be lengthy and therefore consequential for the economics of fusion power plants.
The maintenance strategy that maximizes the economic value of a plant depends on internal factors such as the cost and durability of the replaceable components, the frequency and duration of the maintenance blocks, and the external factors of the electricity system in which the plant operates.
This paper examines the value of fusion power plants with various maintenance properties in a decarbonized United States Eastern Interconnection circa 2050.
Seasonal variations in electricity supply and demand mean that certain times of year, particularly spring to early summer, are best for scheduled maintenance.
Seasonality has two important consequences.
First, the value of a plant can be 15\% higher than what one would naively expect if value were directly proportional to its availability.
Second, in some cases, replacing fractions of a component in shorter maintenance blocks spread over multiple years is better than replacing it all at once during a longer outage, even through the overall availability of the plant is lower in the former scenario.
\end{abstract}

\section{Introduction}

Fusion power plants are likely to have significant periods of scheduled maintenance to replace components in the interior of the reactor, such as the first wall, blankets, and divertors. 
In order to optimize the design of fusion plants and estimate their potential to contribute to future electricity systems, it is important to quantify how the maintenance needs of different plant designs affect their value in an electricity market.
This paper builds on earlier work\cite{schwartz}, which examined the value of fusion power plants in the context of a fully decarbonized power system, namely the United States portion of the Eastern Interconnection.
One conclusion was that the most important characteristics of a fusion plant are its total capital cost and the operations and maintenance cost, particularly the costs of replacing major interior components during scheduled maintenance.
Even if a fusion power plant were to have a net power output that modulates on an hourly timescale (e.g. a pulsed tokamak\cite{biel_systems_2017}), modulation pattern does not affect the marginal value of a plant as determined using an electricity systems model with hourly resolution; only the time-averaged net output of the plant is important.
However, the earlier work did not explicitly model scheduled maintenance periods, and it considered the cost of replacing interior components such as the blanket and divertor during scheduled maintenance periods as \textit{variable} operations and maintenance (VO\&M) costs.
This work, in contrast, assumes that the plant operators are scheduling component replacements far in advance, in some cycle that repeats over an integer number of years.
In this case, replacement costs should instead be considered as \textit{fixed} operations and maintenance (FO\&M) costs.
This paper examines changes to the marginal value of plants when scheduled maintenance is modeled explicitly, at three levels of complexity: first, considering the loss of generation during maintenance periods of various durations and frequencies, second, incorporating a parasitic power draw during maintenance, and third, modeling the finite durability and cost of the components that need maintenance.\vspace{1em}

We find that the loss of value due to maintenance needs is often much smaller than the naive estimate of value being directly proportional to a plant's average annual availability---the fraction of time that it is not under maintenance.
This can inform fusion plant designers by quantifying part of the financial incentive to design a plant to permit shorter duration, or more frequent, scheduled maintenance periods.

\subsection{Maintenance must be considered in fusion electricity-systems studies}
\subsubsection{\textit{Maintenance is especially important for fusion}}

Many fusion plant designs call for regular scheduled maintenance to replace internal components, in particular, the first wall, blankets, and divertors.
These maintenance periods are significantly longer than for other types of power plants.
For example, a 2009 conceptual design study for the European DEMO reactor estimated\cite{nagy_demo_2009} that the plant would require 70 days to replace the divertors after 2 full-power years of operation.
A study on an updated version of the concept in 2014\cite{crofts_maintenance_2014} estimated 8 months for blanket (and divertor) replacements after 4 full-power years of operation.
This leads to relatively low estimates for the average availability of plants, often 75\%\cite{crofts_maintenance_2014, maisonnier_conceptual_2005}
to 85\%\cite{najmabadi_aries-cs_2008}.
For comparison, the United States fission fleet, members of which typically require 32 to 40 days of maintenance every 18 months to 2 years, achieves an overall availability of roughly 90\%\cite{eiafission}. \vspace{1em}

It is therefore important to understand how scheduled maintenance impacts the value of fusion plants in order to both accurately value these plants and to optimally design the overall plant, select components, and plan operating strategies to maximize value.

\subsubsection{\textit{Neglecting variation of electricity prices undervalues a plant}}
\begin{figure}[htbp]
\centering
\includegraphics[width=0.8\textwidth]{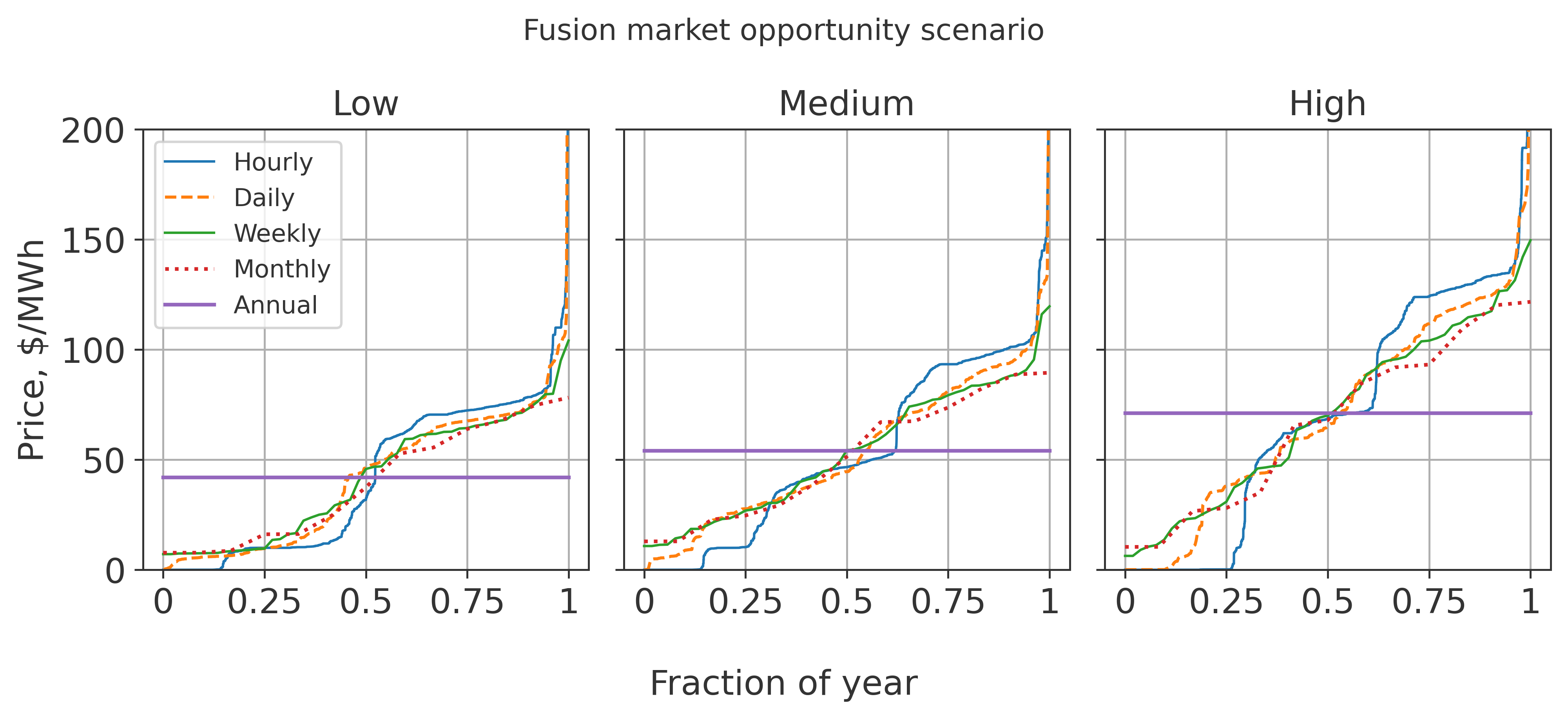}
\caption{Electricity prices averaged in hourly, daily, weekly, monthly, and annual intervals, then sorted, for base cases in the three market opportunity scenarios. Scheduled maintenance can take place when prices are low.
}
\label{fig:pricetimeseriesstructures}
\end{figure}

First, a given plant is likely undervalued if scheduled maintenance patterns are not correctly considered.
Studies of fusion plant conceptual designs usually calculate a levelized cost of electricity (LCOE) that factors in the plant's annual average availability (or its capacity factor).
This LCOE is then often compared to the average electricity price in a country or region, in order to determine whether a plant could be profitable there.
However, this calculation assumes that electricity is equally valuable at all points in time.
In future decarbonized electricity systems, the wholesale price of electricity is likely to exhibit significant seasonal variation
due to variability in supply and demand, e.g. increased solar availability around the summer solstice and electrified heating and air conditioning demands in winter and summer.
Figure~\ref{fig:pricetimeseriesstructures} shows sorted average hourly, daily, weekly, and monthly electricity prices, along with the annual average price, in a particular geographic region\footnote{The PJM-Midatlantic region, generally where the most fusion plants are built in this scenario.} in the three fusion ``market opportunity'' scenarios studied in our earlier work\cite{schwartz}.
Even on a monthly timescale the average price of electricity can be significantly different from the annual average.
Therefore the lost revenue of a fusion plant due to scheduled maintenance depends strongly on the time of year in which the maintenance occurs, not only the duration of the outage events.

\subsubsection{\textit{Correct costs and values are needed for optimization}}
Second, we note that maintenance decisions affect the economics of a fusion plant in multiple ways: maintenance scheduling will affect the plant's value, and the design of replacement components and maintenance equipment to meet a given schedule will affect its cost.
Rather than requiring components to last for several full-power years, in principle a plant could be designed with components which last a few full-power months and are exchanged annually.
This could lead to lower costs for the components themselves (including less lengthy material qualification testing\cite{ibarra_european_2019}) or smaller, higher-neutron-flux devices\cite{sorbom_arc_2015}, but would require more frequent component purchases.
Maintenance downtime would occur more often, but each individual instance could be shorter. \vspace{1em}

Additionally, plant designers may consider the expense of additional equipment if it permits shorter maintenance durations.
For example, Crofts\cite{crofts_maintenance_2014} estimates the duration to replace the plasma-facing components in a DEMO tokamak as a function of the number of remote-handling systems: from 22 months with a single system to 3.2 months with 8 systems.
Alternately, it may be possible to engineer systems to permit quicker access after halting power production,
by choosing lower-activation in-vessel components\cite{seki_impact_1998}.
Designs which permit shorter times before and after component replacement may be especially valuable: if it were feasible to replace a fraction of the components every year rather than all components every few years, the plant could more easily have maintenance scheduled while electricity prices are forecasted to be low.
Finally, shorter but more frequent scheduled maintenance may allow \textit{smaller} numbers of maintenance facilities.
Many fusion plant concepts call for components to be serviced in ``hot cells'' within a maintenance building\cite{iter_hot_cell_2017}.
These heavily-shielded robotic workstations are large and costly, and maintenance strategies that involve replacing a fraction of the interior components at a time, rather than all at once, could require fewer hot cells.
\vspace{1em}

Without a particular plant design it is not possible to evaluate the costs of additional equipment or technology to permit faster access and recommissioning.
This paper, instead, examines the change in the marginal value of a plant as part of the electricity system as a function of the duration and frequency of its required maintenance. These findings can then inform optimal design of future fusion power plants, components, and maintenance regimes.
\vspace{1em}

The rest of this paper is structured as follows: Section~2 studies the loss of value (eg revenue) due to the maintenance outage itself.
Section~3 adds a parasitic power draw during the maintenance period.
Section~4 considers blankets with a finite cost and finite durability against neutron exposure, and determines the optimum replacement strategy.

\section{Value loss due to the maintenance period}
\subsection{\textit {Model and methodology}}
In this work, the open-source electricity system capacity expansion and operations model GenX\cite{genx_software} was used to model a decarbonized future United States Eastern Interconnection electric grid in the year 2050.
The electricity system data and methodology are identical to those in our earlier work\cite{schwartz}, except that fusion and fission plants undergo scheduled maintenance, during which time they produce no power.
While fusion plants require varying amounts of maintenance, fission plants always require a block of 5 weeks every 2 years.
\footnote{These are less frequent, but longer, than their typical outages every 18 months in the present-day United States; fractional-year maintenance cycles have not been implemented in GenX.}\vspace{1em}

During maintenance, a plant neither produces electricity nor requires electric power for its own systems.
A constraint ensures that the total net (nameplate) fusion capacity in the system is \SI{100}{\giga\watt}.
This is a level similar to the present-day deployment of fission in the United States.
For comparison, the average and peak demands in this circa 2050 Eastern Interconnection system are \SI{600}{\giga\watt} and \SI{1100}{\giga\watt}, respectively.
While the previous work examined three scenarios for the costs of resources other than fusion, termed the three ``market opportunity'' scenarios, this study uses only the ``medium'' scenario. Monetary quantities are denoted in 2019 dollars. In general, costs are from the 2021 Annual Technology Baseline. See the previous work\cite{schwartz} and its supplementary material for details. \vspace{1em}

As our prior work determined that pulsed vs. steady state operations had little impact on the marginal value of fusion power plants, plants considered in this study have simpler hourly operations than those in the previous work, and one can imagine them to be a steady-state tokamak, stellarator, or similar, rather than a pulsed tokamak.
The plants are extensions of standard thermal generators with linearized unit commitment.
They have two kinds of parasitic ``recirculating'' power: an \textit{active parasitic power}, equal to 5\% of the gross electrical capacity of the plant, which is drawn from the grid whenever the plant is producing power, and a \textit{passive parasitic power}, equal to 10\% of the plant's gross capacity, drawn at all times except when the plant is undergoing maintenance.
The active parasitic power represents systems such as plasma heating and magnets for plasma control, and pumps for coolant circulation, while the passive parasitic power represents  cryogenics (for superconducting magnets), vacuum pumps, tritium-handling systems, and other house loads.
The plants have a variable operations and maintenance (VO\&M) cost associated with the generator of \$1.74/MWh$_\mathrm{gross}$, but no variable costs associated with the replaceable components such as the blanket; costs for these are modeled as fixed costs, and introduced in Section 4.
Because of the parasitic power, 1 unit of gross capacity corresponds to 0.85 units of net capacity, so the net VO\&M cost is \$2.07/MWh. \vspace{1em}

While in an operational state (i.e. not undergoing maintenance) the plasma and power generation can be turned on and off, with minimum up-times and down-times of six hours.
While generating power, the minimum power level is set at 0.7, and the ramp rate is sufficient to change the power level from 0.7 to 1.0 on an hourly basis.
Unlike in the previous study, there is no significant parasitic load associated with turning the plasma on; this is assumed to be handled by energy storage on-site.\vspace{1em}

Since the cost of a fusion plant as a whole is unknown, we use a methodology (described in more detail in the previous work) to calculate the marginal value of a unit of fusion capacity at a capacity penetration of \SI{100}{\giga\watt}.
The marginal value is the amount by which the cost of the electricity system as a whole would decrease if a marginal (differential) amount of fusion plants could be built for zero capital cost.
It is equivalently, the cost ceiling that actors in the system should be willing to pay for an additional unit of fusion capacity.
It is therefore the cost that a plant of the given design would need to attain in order to reach the given capacity penetration in a long-run equilibrium. \vspace{1em}

Throughout this paper, 
we consider maintenance for a single set of components, or for sets of components which are on identical schedules; i.e. we do not consider plants where blankets and divertors have different replacement schedules.
For brevity, and since it often represents the largest replaceable part of a magnetic fusion device, we generically use the term blanket to signify the replaceable components.
We assume that a given set of components can be replaced all at once, or in a smaller portion each year.
For example, one could replace 1/3 of the blanket during three years of a five year overall cycle. 
We term this a 3:2 \textit{strategy}.
Strategies are denoted by a pair of integers $m:n$ where $m$ is the number of years in the cycle that include a maintenance period, and $n$ is the number of maintenance-free years during the cycle.
The total cycle length $L_\mathrm{cyc}$ is therefore $m + n$.
In this formulation, $L_\mathrm{cyc}$ must be an integer number of years.
Strategies range from $1:L_\mathrm{cyc}-1$, a single block of maintenance, to $L_\mathrm{cyc}:0$, a block of maintenance every year.
The formulation is agnostic to the ordering of maintenance years and maintenance-free years within a cycle.
All maintenance blocks in a cycle are assumed to have the same duration, and for a given unit, are scheduled at the same time of year; for example a plant cannot schedule maintenance in May in two years and October in a third year.
However, because we use a linearized formulation, this is equivalent to $2/3$ of a plant always scheduling maintenance in May and $1/3$ of a plant always scheduling maintenance in October.
The resulting schedule of the number of plants in a given zone undergoing maintenance is equivalent, so allowing plants separate decision variables to schedule each of their maintenance periods would only increase the complexity of the problem, and not further minimize the objective.\vspace{1em}

The duration of each block can be up to a full year.
In the current implementation it is not possible to require more than one maintenance block in a single year, such as changing the whole blanket every six months, or for a maintenance block to be longer than a year.
\vspace{1em}

In the first study, we examine the loss of value associated with maintenance cycles over one to five years, with one or more maintenance blocks per cycle.
For each maintenance strategy considered we examine durations from zero to a full year.
For simplicity, these cases do not include any cost for replacing the blanket, nor any constraints reflecting a finite blanket durability with respect to neutron damage, nor any parasitic power during the maintenance period.
As such, this section estimates only the loss of market revenue or value associated with various maintenance outage regimes, rather than the cost of maintenance and component replacement (which is considered further in Section~4) 

\subsection{\textit {{Results - value loss}}}

\begin{figure}[htbp]
\centering
\includegraphics[width=0.99\textwidth]{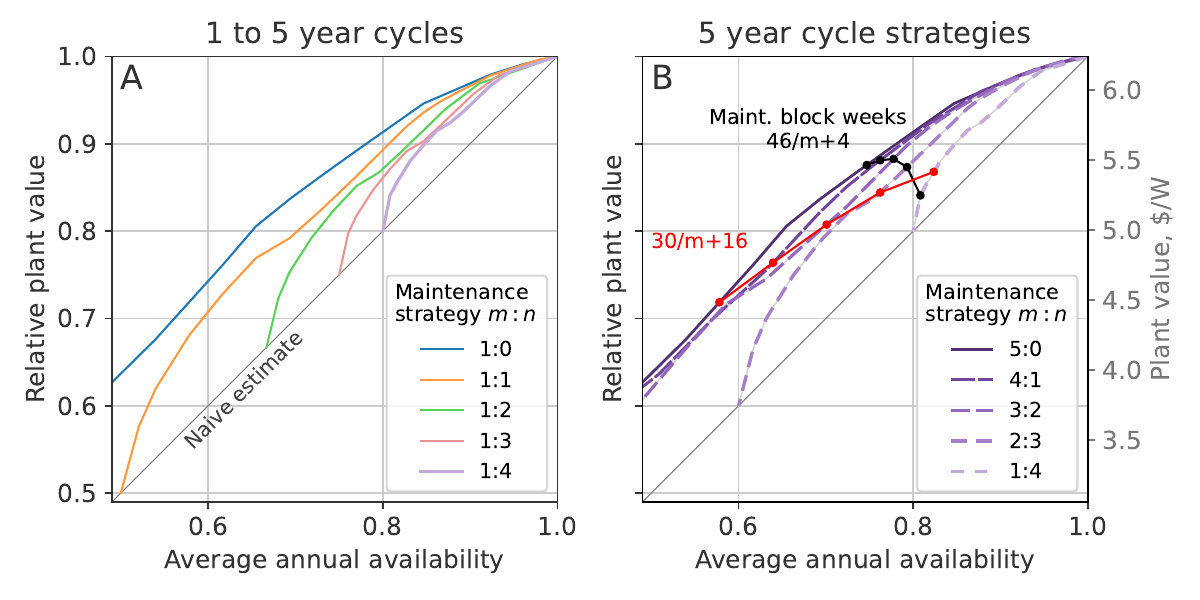}
\caption{Value of plants with the simple maintenance formulation, as functions of their average annual availability.
Part A: plants with one maintenance block per cycle of $m + n = 1$ to $5$ years.
Part B: the value of plants with the five different five-year cycles, from maintenance every year (5:0) to once per five years (1:4).
In this simple formulation, plant values are greater than naive estimate of value being proportional to availability.
Also shown in Part B are the five-year cycle options for two plants: one which requires 46 weeks to exchange interior components plus 4 weeks at the start and end of every maintenance block, and one which requires 30 weeks for exchange, plus 16 weeks at the start and end. 
}
\label{fig:medoppmidrangemaint}
\end{figure}

Fig.~\ref{fig:medoppmidrangemaint} shows the relative marginal value of fusion plants as a function of maintenance strategy and average annual availability.
The average annual availability is $(m (1 - d/8760) + n) / (m + n)$, where $d$ is the duration of the maintenance block in hours.
The relative marginal value is the marginal value compared to that of a plant which requires no maintenance.
Part (A) shows the relative marginal value for cycle lengths of one to five years with a single maintenance block, $m=1$. Part (B) of Figure~\ref{fig:medoppmidrangemaint} shows the relative value of different five-year strategies, as a function of their availability.\footnote{Note that the 1:4 curve in Fig.~\ref{fig:medoppmidrangemaint} (B) is reproduced from Fig.~\ref{fig:medoppmidrangemaint} (A), and the 5:0 curve is identical to the 1:0 curve in Part (A), as both represent strategies with maintenance every year.}
Gray diagonal lines show the naive estimate of the relative marginal value, in which marginal value is directly proportional to availability.
The naive estimate is correct if maintenance outages occur during a period when the average cost of electricity equals the annual average.
In particular this is true for plants requiring a full-year maintenance outage, since
GenX models a single weather year.
Thus, both ends of the curves in these plots must meet the naive estimate. \vspace{1em}

For maintenance periods of less than a full year, the marginal value is significantly higher than the naive estimate. This is because the price of electricity varies throughout the year, and maintenance is scheduled when prices are lowest (as discussed further in Section~\ref{sssec:seasonality}).
For example, a plant with 80\% average annual availability with maintenance every year ($m$:0) retains 91\% of the value of a maintenance-free plant, and even at 65\% availability, its marginal value is 80\% of that of a maintenance-free plant.
For a fixed average availability, the increase over the naive estimate is smaller when there are longer, infrequent maintenance periods: a similar plant requiring 0.6 years of maintenance every three years (in a 1:2 strategy) also has an average availability of 80\%, but retains just 87\% of its value.
A plant requiring a full year of maintenance at a time in a five-year cycle (1:4) has a relative value equal to its average availability of 80\%. These results indicate that shorter but more frequent maintenance periods retain more value than longer, less frequent outages resulting in the same average availability.

\subsection{Choices between strategies}
In an operating fusion plant there may be multiple possible strategies for replacing components.
All replaceable components could be exchanged in one longer scheduled block, or a fraction of them could be replaced in multiple shorter blocks.
The optimal maintenance strategy will depend on properties of the components to be replaced, such as their cost and working lifetime, and properties of the plant as a whole.
In this section we model a given plant as having:
\begin{itemize}
\item a duration to shut down the plant, prepare for replacement activities, finalize replacement activities, and recommission the plant, called the maintenance start-stop time, $T_\mathrm{ss}$;
\item and a duration to exchange a whole blanket, $T_\mathrm{exch}$, which does not include the start-stop time.
\end{itemize}
We assume that the blanket is composed of many sub-components, and that once replacement activities are underway, exchanging a fraction of the blanket takes that fraction of the whole-blanket replacement time.
Thus, a single maintenance block to replace the whole blanket would require $T_\mathrm{exch} + T_\mathrm{ss}$, and a block to replace half the blanket would require $T_\mathrm{exch}/2 + T_\mathrm{ss}.$
The cost and working lifetime of the replaced components are ignored until Section~4. \vspace{1em}

Part (B) of Figure~\ref{fig:medoppmidrangemaint} illustrates the effect of maintenance strategy choice, within a given cycle length $L_\mathrm{cyc} = 5$ years, for two illustrative example plants with different resulting behaviors.
One plant has $T_\mathrm{exch} = 46$ weeks and $T_\mathrm{ss} = 4$ weeks.
The different possible strategies for this plant are shown in black.
The points for each strategy must lie on the corresponding curve.
The 3:2 strategy, with three 19$\frac{1}{3}$-week maintenance blocks, has the highest value, which is 4\% higher than that of the 1:4 strategy with a single 50-week block, even though the average annual availability is 3\% lower.
Another plant, shown in red, has $T_\mathrm{exch}$ and $T_\mathrm{ss}$ of 30 and 16 weeks, respectively.
The 1:4 strategy with a single block of 46 weeks is value-optimal since the start-stop time is relatively long.

\subsubsection{\textit {{Seasonal maintenance cycles}}} \label{sssec:seasonality}
\begin{figure}[htbp]
\centering
\includegraphics[width=0.85\textwidth]{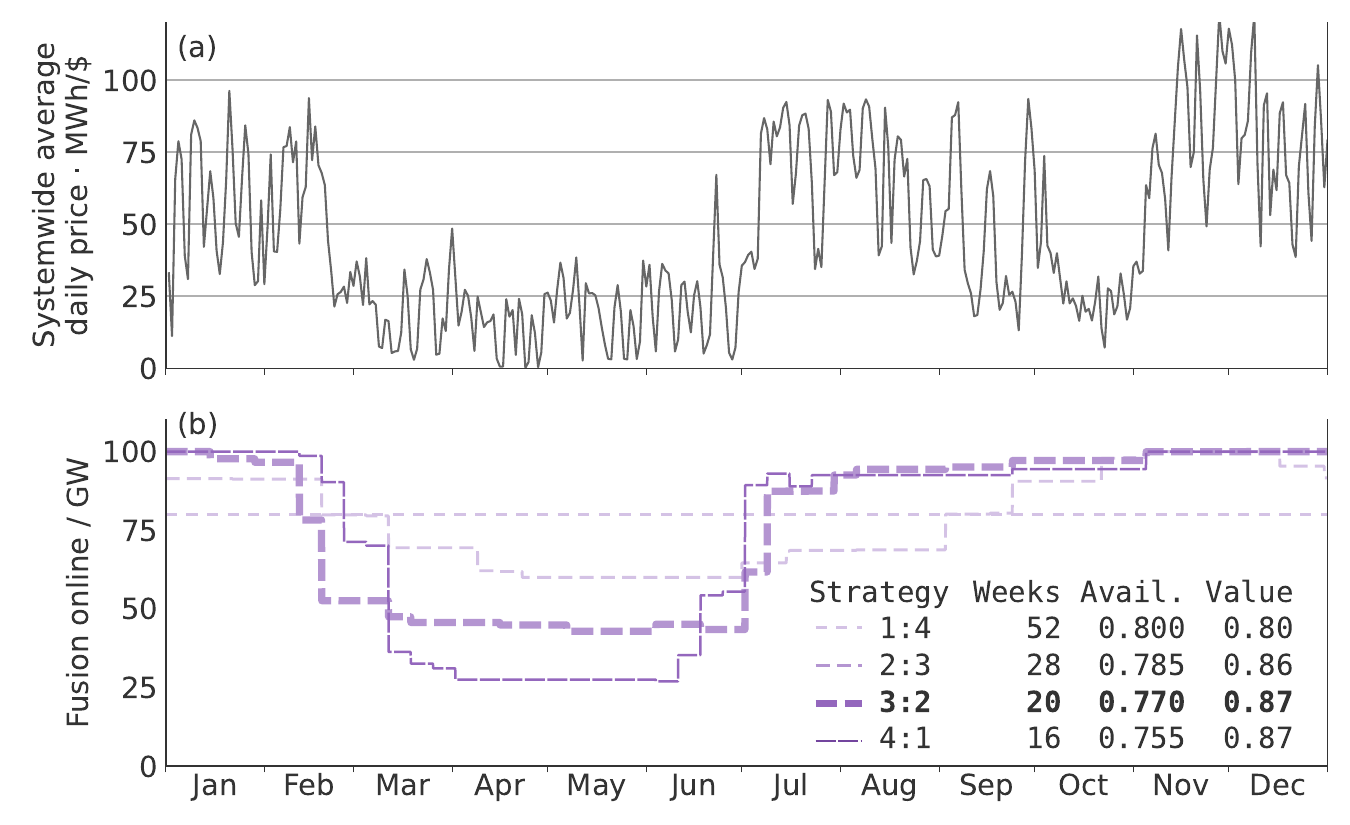}
\caption{Part (a): daily average electricity prices, averaged across the 20 modeled geographic zones. Part (b): capacity of fusion plants online each day of the year under optimized maintenance schedules, for four of the five-year strategies for a plant with $T_\mathrm{exch} = 48$ weeks and $T_\mathrm{ss} = 4$ weeks. The highest value strategy (3:2), with three 20-week maintenance blocks over the five year cycle, is not the one with the highest availability. `Value' is relative to the marginal value of a maintenance-free plant.}
\label{fig:maintts}

\end{figure}

Plants generally have a higher value than would be implied by a naive estimate from their availability, because maintenance can be scheduled at the time of year when electricity prices are lowest. Fig.~\ref{fig:maintts} (a) shows the daily system-wide average electricity price in a reference case without any fusion plants. (Note that this average is not weighted by demand levels between different zones.) Prices are significantly lower in spring to early summer than at other times of year, as renewables are abundant and heating and cooling loads are low. 
Fig.~\ref{fig:maintts} (b) shows the three optimized yearly maintenance schedules for plants with $T_\mathrm{exch}$ and $T_\mathrm{ss}$ of 48 and 4 weeks, respectively, with $L_\mathrm{cyc} = 5$ years.
The 1:4 strategy requires a full-year maintenance block, so 20 GW of fusion capacity is under maintenance at all times and value is reduced by 20\% compared to a maintenance-free design.
The 2:3 strategy requires a 28-week maintenance block.
Even though the average annual availability is slightly lower than that of the 1:4 strategy, at 78.5\%, the ability to schedule more of the maintenance periods in the spring, and have all plants online during the period of highest prices in November-December, means that the relative marginal value of these plants is higher, at 86\% of the value of a maintenance-free plant.
The 3:2 strategy, with 20-week maintenance blocks, schedules almost all of the downtime in the lowest price periods, allowing a value of 87\% that of the maintenance-free plant even though the average annual availability is 77\%.
While in this optimization, which includes perfect foresight for weather conditions, the 4:1 strategy also has a relative value of 87\%, in a real system its shorter 16-week maintenance periods could give operators more flexibility or contingency to complete maintenance during the period of lowest prices.

\subsection{\textit{Comparison of results to simple model}}
It is useful to approximate the effect of different maintenance strategies on the value of a plant in a given electricity system without running a full optimization.
This section demonstrates that this may be reasonable, especially for plants with longer $L_\mathrm{cyc}$ and higher availability. \vspace{1em}

\begin{figure}[htbp]
\centering
\includegraphics[width=0.50\textwidth]{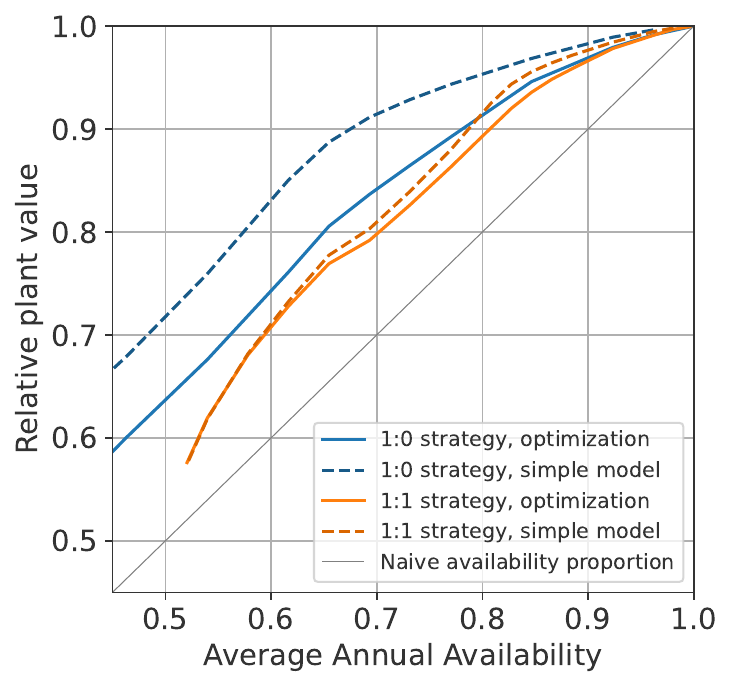}
\caption{Relative marginal value approximated with Eq.~\ref{eq:simplemodel}, compared to the optimized-maintenance results, for plants with 1:0 and 1:1 strategies.
The two optimized curves are the same as those in Figure~\ref{fig:medoppmidrangemaint}.}
\label{fig:threemodels}
\end{figure}

One can approximate the loss of marginal value for a plant with a given maintenance duration by analyzing price series from a case \textit{without} explicitly-modeled maintenance.
Fig.~\ref{fig:threemodels} compares an approximation to the results of the full model.
Here, the relative marginal value of a plant in a given zone is approximated by assuming no revenue during the lowest-average-price period,
\begin{equation}\label{eq:simplemodel}
1 - \frac{\min_s \sum_{t = s}^{s+d} p(\text{mod}_{8760}(t))}{\sum_{t=1}^{8760} p(t)},
\end{equation}
where $p(t)$ is the price in hour $t$, $s$ is the start of maintenance, and $d$ is the maintenance duration.\vspace{1em}

For a case with a 1:0 maintenance strategy, the model approximates the results of the optimized maintenance formulation, though it overestimates the relative value by up to 10\%.
This simple model overestimates the plant's value because it assumes that maintenance for all plants in a zone can be scheduled at the same time, and that this has no effect on the supply (and resulting price) of electricity. 
The model is significantly better for plants with lengthier maintenance cycles: for a 1:1 strategy it overestimates the relative value by only 2.5\% at most; the result for 1:2 strategies (not shown) is similar.\vspace{1em}
We expect that the deviation from this simple model would be greater in systems with a larger fraction of power supplied by fusion or other plants requiring long maintenance periods, or in systems where the discrete nature of individual units is important.
However, this result suggests that for systems with only a few fusion plants or where the plants have longer cycles and high availability, it is not necessary to run a detailed power grid optimization to estimate the value of a given maintenance strategy. 

\section{Parasitic power during maintenance}
A fusion plant may require a significant parasitic power load even during the maintenance period, to continue to run cryogenic, vacuum, ventilation, or other systems\cite{kerekes_operational_2023}.
Depending on which systems must still run, the loads may be comparable to those during operations, especially to those present in a ``warm'' state while the plasma is off, here termed the ``passive'' parasitic load.
In this section the model accounts for these loads.
We examine cases where the fraction remaining during maintenance is 0, 50\%, or 100\% of the passive parasitic load during operations.
The plants in this study have a passive parasitic load equal to 10\% of their gross power output or 11.7\% of their net power output. 
\vspace{1em}

\begin{figure}[htbp]
\centering
\includegraphics[width=0.5\textwidth]{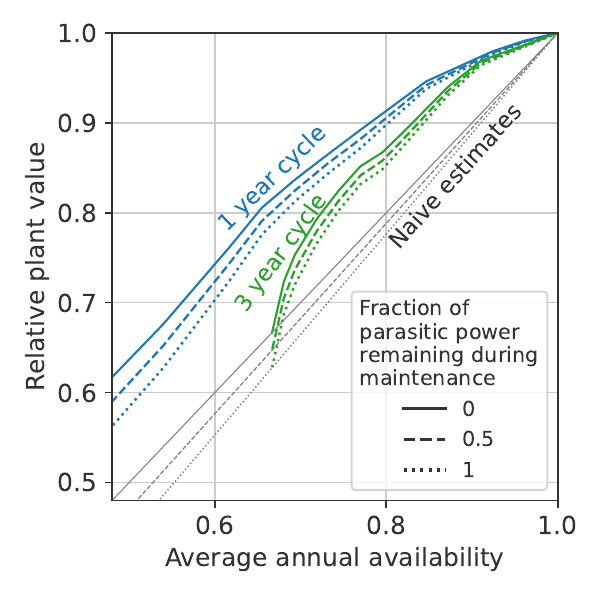}
\caption{Marginal values of plants when they draw parasitic power during maintenance. During the operating period all plants have a parasitic power equal to 10\% of gross capacity. Dashed and dotted lines depict where 50\% and 100\% of this, respectively, continues during the maintenance period.
The losses of value (differences between curves of the same color) are comparable to the naive estimates (differences between gray lines), and decrease at high availability.}
\label{fig:parasiticpassive}
\end{figure}

Figure~\ref{fig:parasiticpassive} shows the relative marginal values of plants with 1-year and 3-year maintenance cycles as a function of their availability.
The gray ``naive estimate'' lines run between $(1,1)$ and $(0, -0.1 * f_\mathrm{rdm}/0.85)$, where $0.1$ is the passive parasitic power fraction during operations, $f_\mathrm{rdm} = \{0, 0.5, 1\}$ is the fraction of that which remains during maintenance, and $0.85$ is the ratio of net to gross generation for these plants.
The loss due to parasitic power during maintenance (differences between curves of the same color) is slightly smaller than one would naively estimate, especially at high availability.
For example, example at an availability of 0.9, one would estimate that plants with all of their passive parasitic power remaining during maintenance should have a relative value 1.2\% lower than a plant without any parasitic power during maintenance. However, only 0.4\% and 0.6\% are lost for plants with 1-year and 3-year cycles, respectively.
The loss becomes closer to the simple estimate as availability decreases: at 0.7 availability, plants with 1- and 3-year cycles have relative values 2.4\% and 3.2\% lower, respectively, than plants with $f_\mathrm{rdm}=0$; compare to the naive estimate for these plants of a 3.5\% lower relative value.
In general, the penalty for parasitic power during maintenance is smaller than the naive estimate because maintenance periods are scheduled when electricity prices are low.

In the next section, plants are assumed to require during maintenance half of the parasitic power draw during operation, or 5\% of the gross capacity.

\section{Optimal maintenance strategies with explicit blanket lifetimes and costs}
This section introduces costs and finite durability for the components that need replacement, in addition to the existing whole-blanket exchange time $T_\mathrm{exch}$ and the maintenance start-stop time $T_\mathrm{ss}$.
This allows us to directly compare maintenance cycles with different $L_\mathrm{cyc}$, and allows for more sophisticated comparison between different strategies with the same $L_\mathrm{cyc}$. \vspace{1em}

Interior components of fusion reactors that need replacement generally do so because they are degraded by exposure to neutrons (and more generally,  subsequent forms of radiation, such as gamma rays) produced by the fusion reactions, or by exposure to the plasma itself.
Degradation due to radiation is essentially proportional to the number of fusion reactions that have occurred in the device, and therefore in a working power plant, closely proportional to the gross electric energy production.
The same is not true for damage to plasma-facing components, such as from erosion, which is highly nonlinear and dependent on plasma conditions.
One might reasonably assume that for plasma operation close to the nominal 100\% power condition, damage is proportional to time with plasma operations.
Depending on the design, there is potential for damage due to cyclic thermal fatigue, from changing the operating power level or turning the plasma on and off.
However, for simplicity we model all degradation as exactly proportional to gross electric energy production.
The lifetime of components in a plant is often quoted as being a certain number of full-power years (FPY), meaning the time they would last if the plant were to operate continuously at its nominal power. \vspace{1em}

Because interior components are expensive and maintenance is a lengthy process, plant designers and operators may need to choose between operating plants at a high duty factor and replacing components more often, or operating plants only when they are most needed (i.e. when electricity prices are relatively high) and replacing the components over a greater number of years, thus lowering the annual replacement cost. Here we quantify these tradeoffs and identify optimal maintenance strategies for various component cost and durability scenarios.

\subsection{Cases and methodology}
This section uses the same electricity system and plant design as in Section~3, but adds to the plant model a cost per full blanket and a durability in full-power years (FPY).
Plants with various combinations of maintenance durations $(T_\mathrm{exch} ,T_{ss})$ are examined.
For each of these combinations, and for blanket costs and durability from \$0 to \$500M/GW$_\mathrm{gross}$ and 0.7~FPY to 5~FPY, respectively, we select the best maintenance strategy $m:n$ out of the 21 with $L_\mathrm{cyc}$ of up to 6 years.
Blanket durability is scanned in intervals of 0.1~FPY, and blanket costs are examined in intervals of \$10M/GW.
If all combinations were performed as separate simulations this would require $51 \times 42 \times 21$ simulations per pair of $(T_\mathrm{exch} ,T_{ss})$, but in practice the set of simulations needed is much smaller.
First, for a given maintenance strategy and durability, the sum of annual blanket cost and the marginal value that the plant provides to the system annually is a constant; one can trivially trade off the two (e.g., a plant with higher maintenance cost has equivalently lower marginal value).
Therefore simulations can be performed with a blanket cost of 0, and plant values at finite blanket cost can be calculated.
Second, one can calculate a maximum useful durability for a blanket with a given $L_\mathrm{cyc}$ and strategy: for example, a plant with a blanket changed every year gains no benefit from increasing its durability beyond $(\mathrm{1\;year} - T_\mathrm{exch} - T_\mathrm{ss})$ since the durability constraint is no longer binding.
Third, one can exploit the structure of the problem:
the region where a given cycle length or strategy dominates is contiguous, increasing durability for a given $L_\mathrm{cyc}$ favors higher $n$ over lower $n$, increasing cost favors lower $L_\mathrm{cyc}$, and increasing durability favors higher $L_\mathrm{cyc}$.

\vspace{1em}

We first discuss the effect of changing the blanket cost, in isolation, and the durability, in isolation, for a plant design with $T_\mathrm{exch}$ of 10 weeks and $T_\mathrm{ss}$ of 4 weeks.

\subsection{Results}

\begin{figure}[htbp]
\centering
\includegraphics[width=0.85\textwidth]{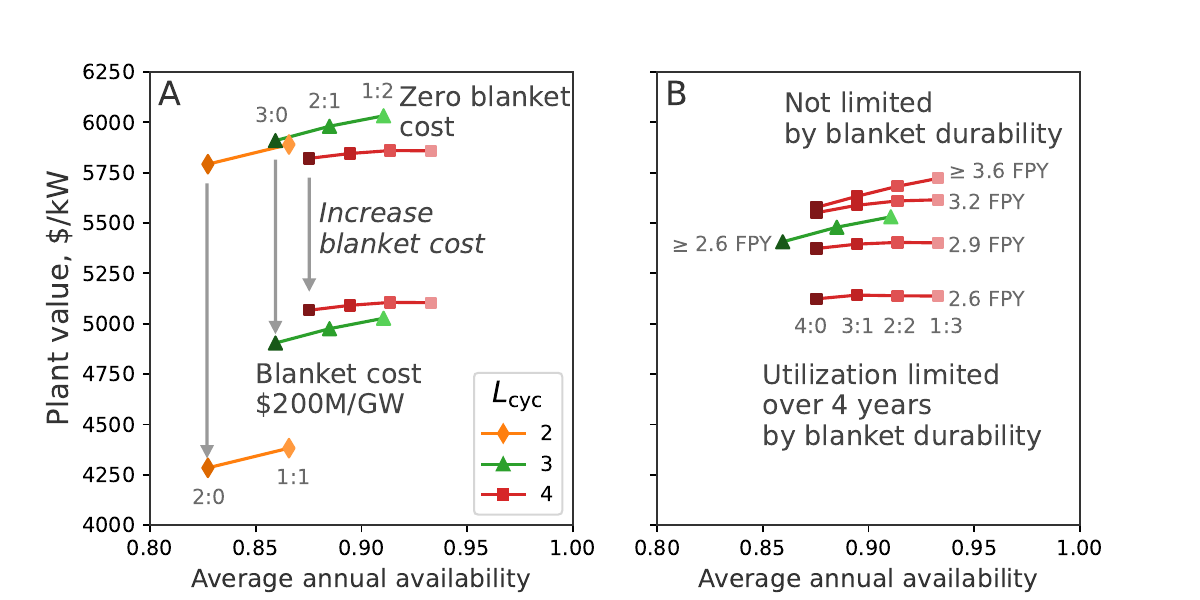}
\caption{Changes in the optimal maintenance strategy with varying blanket cost and durability, $T_\mathrm{exch}$ of 10 weeks and $T_\mathrm{ss}$ of 4 weeks. Lines join groups with the same cost, durability, and $L_\mathrm{cyc}$. Small labels show the strategies m:n and in Part B, varying durability in~FPY. Part A shows the effect of increasing the blanket cost from 0 to \$200M/GW$_\textrm{gross}$, while keeping the design lifetime fixed at 3 full-power years (FPY). The optimal strategy changes from a three-year cycle to a four-year cycle. 
Part~B shows how increasing the blanket lifetime at fixed cost changes the optimal cycle.
If the design lifetime can be increased from 2.6~FPY to about 3.1~FPY, the optimal strategy changes from a three-year cycle to a four-year cycle.}
\label{fig:twoeffects}
\end{figure}

\subsubsection{\textit{Finite blanket cost}}
A longer cycle becomes more favorable as blanket costs increase, because spreading the blanket's cost over more years and targeting operation during higher-value hours in those years mitigates the larger fixed cost.
Figure~\ref{fig:twoeffects}, Part~A shows the effect of increasing the blanket cost on the optimal maintenance cycle, for a blanket with a durability of 3~FPY.
Average annual availability and the plant marginal value are shown for all the maintenance strategies with $L_\mathrm{cyc}$ of 2, 3, and 4~years.
As the blanket cost increases, the value of plants with any maintenance strategy decreases by an amount proportional to the cost and inversely proportional to the cycle length: as shown with arrows in Part~A the value of the 2-year strategies decrease twice as fast as the value of the 4-year strategies.
For this example, if blankets had zero cost, a three-year cycle (1:2) has the highest value.
To help explain this, we can examine the utilization factor, which we define as the ratio of the yearly average gross power generation to the gross capacity of the plant.\footnote{This is slightly different from the conventional definition of ``capacity factor'', the ratio between net generation and net capacity.}
Plants on the 1:2 strategy have a utilization factor of 0.86, while the four-year cycle is limited to a utilization factor of $3/4$ because the blanket's durability is 3~FPY.
However, but with a blanket cost of \$200M/GW$_\mathrm{gross}$, the highest-value cycle changes to one with four years. 
The annual blanket cost per year is \$50M/GW$_\mathrm{gross}$/year for a four-year cycle but \$67M/GW$_\mathrm{gross}$/year for a three-year cycle.
The constrained utilization results in an average yearly revenue which is lower by about \$8M/GW$_\mathrm{gross}$/year for plants on a four-year cycle, but this is smaller than the cost advantage of \$17M/GW$_\mathrm{gross}$/year.

\subsubsection{\textit{Finite blanket lifetime}}
Blankets must be replaced because they have a finite lifetime.
However, whether or not this becomes a constraint on day-to-day operations depends on the chosen maintenance strategy. 
For some strategies, the maintenance duration may be long enough that the plant could be run continuously at full power during operational periods.
Other times, it might make sense to `stretch' a blanket over a longer cycle, while ensuring that the accumulated energy generation does not exceed its durability.
Figure~\ref{fig:twoeffects}, Part~B shows the effect of blankets with different durability on value, for plants with $T_\mathrm{exch}$ of 10 weeks, $T_\mathrm{ss}$ of 4 weeks, and a blanket cost of \$100M/GW.
Average annual availability and plant marginal value are shown for all strategies with 3 or 4-year cycles, for various lengths of blanket durability of 2.6~FPY and up.\vspace{1em}

For plants of this maintenance duration on a 3-year cycle (green triangles), 2.6~FPY is sufficient.
While this is slightly lower than the availability of 2.73~years for the 1:2 strategy, the price of electricity is sometimes low enough that the plant reduces its power output, and the average utilization of the plants over the cycle is 2.58~FPY.
On the other hand, for plants on 4 year cycles, which support 3.50 to 3.73 years of operating time,  2.6~FPY of blanket durability (bottom row of red squares) places a significant limit on daily operations.
All the 4-year strategies have similar value, which will result in similar choices of when to generate power.
If the durability can be increased to 3.2~FPY, a four-year cycle becomes favorable over a three-year cycle.
While the utilization is still limited, the plant can prioritize operating at times of high electricity prices, and has a lower annual blanket cost than a three-year cycle.
Increasing the durability further to 3.6~FPY provides additional benefit to the cycles with higher availability (2:2 and 1:3---those furthest to the right), but going beyond that point there is no additional benefit, as even the 1:3 strategy has a maximum of 3.73 years of operating time.
As the next section will show, increasing the durability can alter the choice of best strategy for a blanket with a given cost.

\subsubsection{\textit{Optimal strategies}}
\begin{figure}[htbp]
\centering
\includegraphics[width=0.95\textwidth]{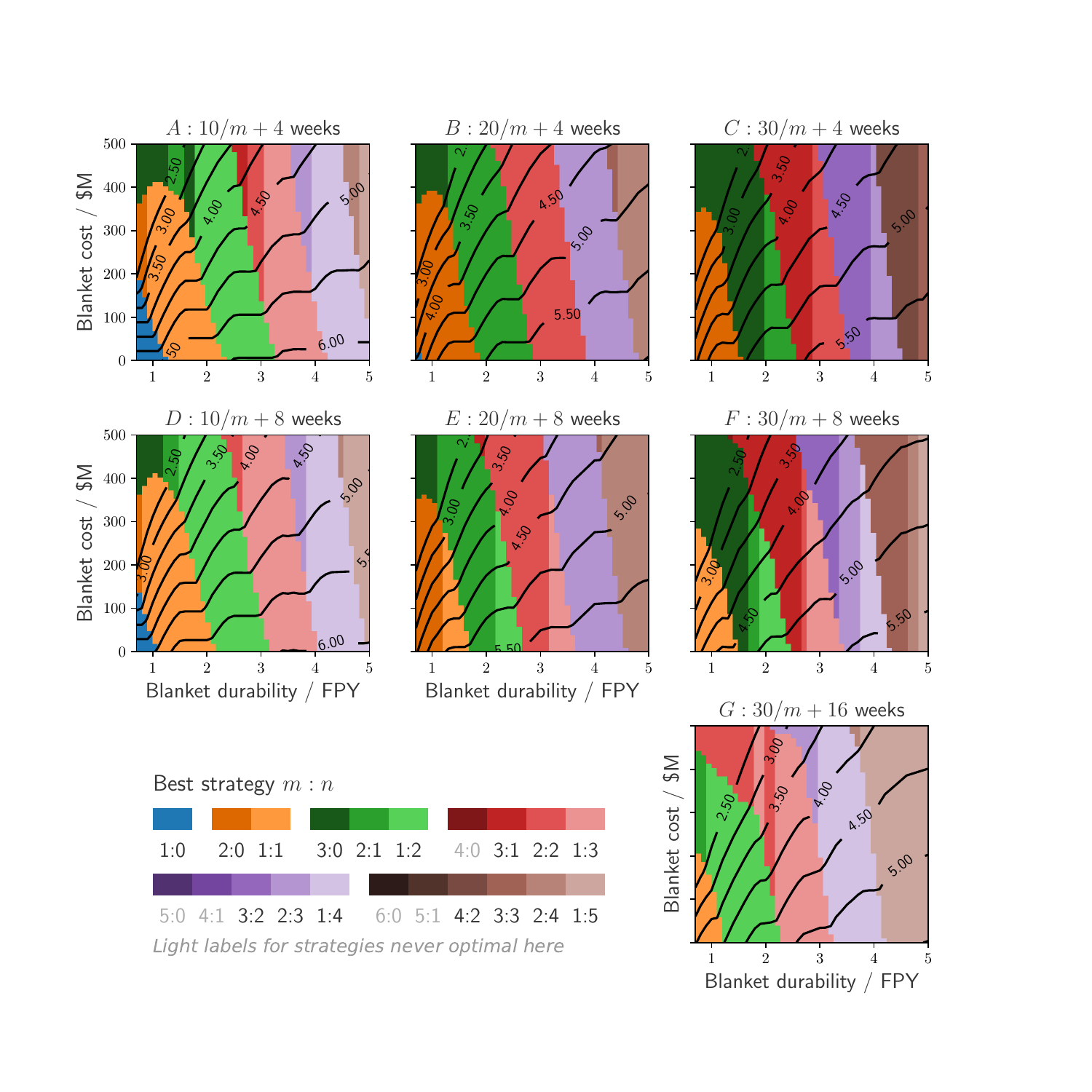}
\caption{Maps of the optimal maintenance strategy and plant value as function of blanket lifetime and cost, for seven sets of maintenance durations $(T_\mathrm{exch}, T_\mathrm{ss})$.
For a given maintenance durations, with a chosen strategy m:n, a maintenance block takes $T_\mathrm{exch}/m + T_\mathrm{ss}$ weeks.
Colors present the optimal strategy, while contours present the marginal value in \$/W.
Low durability and low cost favor blanket replacement more often, while higher costs and higher durability favor replacement less often.
Darker shades within each color are present when the highest-value strategy uses short maintenance blocks during multiple years of the cycle, rather than one longer block in which the whole blanket is replaced.
}
\label{fig:fouroptimal}
\end{figure}
Figure~\ref{fig:fouroptimal} shows the optimal maintenance strategies as functions of the blanket cost and durability (in FPY) for seven sets of maintenance durations $(T_\mathrm{exch}, T_\mathrm{ss})$, as well as a contour plot of the marginal value of the plant at \SI{100}{\giga\watt} of capacity penetration. \vspace{1em}

Let us first remark on the stepped nature of the contours. 
While, naturally, increasing the blanket cost can only decrease the plant's value, 
and increasing blanket durability can never decrease the plant's value, in some cases it does not increase the value either, and the contour is horizontal.
This happens when the blanket's durability constraint is not binding, but it would be if it were operated over a longer cycle. 
For example, in Part~B ($20/m+4$ weeks), with \$100M blanket that lasts 3.5~FPY, the optimal strategy is to replace it with a 2:2 strategy in a four year cycle, which has 3.46~years of possible operational time.
There is no additional value to be gained by increasing the durability to 3.7~FPY while keeping the strategy, since $3.5 > 3.46$. 
If the blanket were operated over a 5 year cycle, the plant's operation would necessarily be restricted to not exceed 3.7~FPY; this turns out to be sub-optimal.
Only at 3.8~FPY does the optimal cycle length increase to 5 years with a 2:3 strategy; at that point marginal increases to durability become valuable again.\vspace{1em}

Now let us discuss the optimal strategies.
Plants with long $T_\mathrm{exch}$ relative to $T_\mathrm{ss}$, such as in part C (top right) favor  multiple maintenance blocks per cycle, shown in darker shades.
With 30 weeks to exchange a whole blanket plus just 4 weeks to start and stop maintenance, the optimal strategy for these plants always involves between two and four maintenance blocks per cycle---for example if the blanket has a durability of more than 4~FPY, there is a region (dark brown, right) where the optimal strategy is 4:2, or four blocks of 11.5~weeks over a six-year cycle.
Decreasing $T_\mathrm{exch}$ (moving left to B and A) tends to favor higher-availability strategies with fewer maintenance blocks per cycle, shown with lighter shades.
Increasing $T_\mathrm{ss}$ (moving down to F and G) has a similar effect of favoring single-maintenance-block strategies, as average availability sharply falls with each additional block. \vspace{1em}

Within the region of a single cycle length (bands of a single color), increasing durability tends to favor fewer maintenance blocks per cycle (lower $m$, lighter shades), as higher-$m$ strategies would not support enough operational time to take advantage of the higher durability.
Note that exceptions to this rule in $A$ and $G$, where darker bands appear to the right of lighter bands of the same color, are likely due to finite precision in the solver---the different strategies have values within about 1\% of each other. \vspace{1em}

Not all possible strategies appear on this plot: even in Part~C the highest-$m$ strategies for $L_\mathrm{cyc}$ of 4, 5, and 6 years are never optimal.
In the extreme case where $T_\mathrm{ss}$ is reduced to zero, the highest-$m$ strategy for a given $L_\mathrm{cyc}$ would always be chosen in this model, as it allows for the most flexibility in scheduling.

\section{Discussion and conclusion}

\subsection{Directions for future work}
At the time of writing, the maintenance model does not capture maintenance cycles of 6 or 18 months, or other non-integer-year cycles. Cycles of an integer-plus-half years, for example, would require constraints that couple the maintenance state at hour $t$ with that half a year away, in hour $t + 4380$.
These could potentially be useful if the blanket design has a short lifetime, but the alternating maintenance season would diminish the benefit that integer-cycle strategies gain from scheduling maintenance in the season with the lowest electricity prices.
Generally, this model could be extended to handle plants that require multiple maintenance periods per year, for example, a plant with short-lived plasma-facing components.
Other plants could have separate maintenance cycles for two or more sets of components, such as blankets and divertors for large magnetic fusion plants, with different outage durations.
\vspace{1em}

This work does not assume any particular extra cost with each maintenance block.
In practice there could be fixed costs (for example, relating to decontamination of maintenance equipment, or consumables) associated with each maintenance block regardless of its duration. 
With estimates for these costs, one could easily re-analyze the data generated in this work to find a new optimal strategy. \vspace{1em}

\subsection{Conclusions}
The value of a plant, and the optimal maintenance strategy, depends on the electricity system it inhabits.
Future mostly- or wholly-decarbonized electricity systems may have a significant seasonality in their wholesale electricity prices; this can be exploited by fusion plants or other generators requiring long maintenance periods.
By scheduling maintenance in seasons with low electricity prices, a loss of value as large as the overall plant unavailability can be avoided; losses can be half of the naive expectation.
This suggests that plant designs need not achieve as high availability as was previously assumed (typically 80\%).
We also find that there is an opportunity for plants to be more profitable with frequent short outages rather than infrequent longer outages.
This also provides new urgency for the development of methods to reduce the time between a shutdown and the start of remote handling operations and methods to accelerate re-commissioning.

\section*{Acknowledgements}
Thanks to Ruaridh Macdonald for discussions that led to an improved modeling approach for maintenance.
J.A.\ Schwartz and E.\ Kolemen were supported through the U.S.\ Department of Energy under contract no.\ DE-AC02-09CH11466.
The United States Government retains a non-exclusive, paid-up, irrevocable, world-wide license to publish or reproduce the published form of this manuscript, or allow others to do so, for United States Government purposes.
J.A. Schwartz and W.\ Ricks also received support for this work from the Princeton Zero-Carbon Technology Consortium, which is supported by unrestricted gifts from General Electric, Google, ClearPath, and Breakthrough Energy.

\section*{Disclosure of Interests}
J.D. Jenkins is part owner of DeSolve, LLC, which provides techno-economic analysis and decision support for clean energy technology ventures and investors. A list of clients can be found at \url{https://www.linkedin.com/in/jessedjenkins}. He serves on the advisory boards of Eavor Technologies~Inc., a closed-loop geothermal technology company, and Rondo Energy, a provider of high-temperature thermal energy storage and industrial decarbonization solutions, and has an equity interest in both companies. He also serves as a technical advisor to MUUS~Climate Partners and Energy Impact Partners (EIP), both investors in early stage climate technology companies. EIP is an investor in Zap~Energy.

\section*{Data and code availability}
Data will be made available at \url{https://doi.org/10.34770/jsyh-gg45}.

\section*{References}

\begin{enumerate}
    \renewcommand{\labelenumi}{[\theenumi]}

    \item SCHWARTZ, J., RICKS, W., KOLEMEN, E., JENKINS, J., The value of fusion energy to a decarbonized United States electric grid, Joule 7 4 (2023) 675--699.

    \item BIEL, W., et al., Systems code studies on the optimization of design parameters for a pulsed DEMO tokamak reactor, Fus.\ Eng. \ Des.\ 123 (2017) 206--211.

    \item NAGY, D., BONNEMASON, J., DEMO divertor maintenance, Fus.\ Eng.\ Des.\ 84 7--11 (2009) 1388--1393.

    \item CROFTS, O., HARMAN, J., Maintenance duration estimate for a DEMO fusion power plant, based
    on the EFDA WP12 pre-conceptual studies, Fus.\ Eng.\ Des.\ 89 (2014) 2383–-2387.

    \item MAISONNIER, D., et al., A conceptual study of commerical fusion power plants: Final report of the European Fusion Power Plant Conceptual Study (PPCS), EFDA-RP-RE-5.0, European Fusion Development Agreement, 2005.

    \item NAJMABADI, F., et al., The ARIES-CS compact stellarator fusion power plant, Fus.\ Sci.\ Tech.\ 54 3 (2008) 655--672.

    \item UNITED STATES OF AMERICA ENERGY INFORMATION ADMINISTRATION, Table 6.07.B.\ Capacity factors for utility scale generators primarily using non-fossil fuels (2023),\\
    \url{https://www.eia.gov/electricity/monthly/epm_table_grapher.php?t=table_6_07_b}
    
    \item IBARRA, A., et al., The European approach to the fusion-like neutron source: the IFMIF-DONES project, Nucl.\ Fusion 59 6 (2019) 065002.

    \item SORBOM, B., et al., ARC: A compact, high-field, fusion nuclear science facility and demonstration power plant with demountable magnets, Fus.\ Eng.\ Des.\ 100 (2015) 378--405.

    \item SEKI, Y.,  et al., Impact of low activation materials on fusion reactor design, J.\ Nucl.\ Mat.\ 258--263 (1998) 1791--1797.

    \item FRICONNEAU, J-P. et al., ITER hot cell---Remote handling system maintenance overview, Fus.\ Eng.\ Des.\ 124 (2017) 673---676.
    
    \item  MIT ENERGY INITIATIVE, PRINCETON UNIVERSITY ZERO LAB. GenX: a configurable power system capacity expansion model for studying low-carbon
    energy futures (2023),\\
    \url{https://energy.mit.edu/genx/}

    \item KEREKE\v{S}, A.,  et al., Operational characterization of tokamak and stellarator type fusion power plants from an energy system perspective, Fus.\ Eng.\ Des.\ 190 (2023) 113496.

\end{enumerate}

\end{document}